# Identifying the Genetic Basis of Functional Protein Evolution Using Reconstructed Ancestors

Victor Hanson-Smith[1], Christopher Baker[2], and Alexander Johnson[1]

*1 Department of Microbiology and Immunology, University of California San Francisco*

*2 Department of Plant and Microbial Biology, University of California Berkeley*

Editorial correspondence to:

Alexander Johnson
N372 Genentech Hall, Mission Bay
600 16th Street, Box 2200
San Francisco, CA 94158
415-476-8783

ajohnson@cgl.ucsf.edu

## Abstract

A central challenge in the study of protein evolution is the identification of historic amino acid sequence changes responsible for creating novel functions observed in present-day proteins. To address this problem, we developed a new method to identify and rank amino acid mutations in ancestral protein sequences according to their function-shifting potential. Our approach scans the changes between two reconstructed ancestral sequences in order to find (1) sites with sequence changes that significantly deviate from our model-based probabilistic expectations, (2) sites that demonstrate extreme changes in mutual information, and (3) sites with extreme gains or losses of information content. By taking the overlaps of these statistical signals, the method accurately identifies cryptic evolutionary patterns that are often not obvious when examining only the conservation of modern-day protein sequences. We validated this method with a training set of previously-discovered function-shifting mutations in three essential protein families in animals and fungi, whose evolutionary histories were the prior subject of systematic molecular biological investigation. Our method identified the known function-shifting mutations in the training set with a very low rate of false positive discovery. Further, our approach significantly outperformed other methods that use variability in evolutionary rates to detect functional loci. The accuracy of our approach indicates it could be a useful tool for generating specific testable hypotheses regarding the acquisition of new functions across a wide range of protein families.

## Significance Statement

This study presents a new method to identify historic amino acid substitutions that were responsible for creating specific protein functions in present-day species. This method provides a powerful hypothesis-generation tool for molecular experimentalists studying the underlying mechanisms of protein function.

## Introduction

Evolutionary variation in proteins underlies a great deal of organismal novelty. However, it is often challenging to identify the specific amino acid substitutions that caused protein functions to diversify in related species. A simple approach to this challenge is to directly compare homologous protein sequences, whose functions differ in contemporary species, in order to reveal their amino acid differences (Fig. 1a). The usefulness of this "horizontal" approach is limited because many of the differences between protein sequences have no effect on their functions; in other words, a horizontal comparison does not provide a systematic means to differentiate the small number of amino acid changes responsible for shifting protein function from the relatively large number of amino acid changes that accumulate over evolutionary time without affecting function. Further, a horizontal approach is blind to directionality of change, that is, which amino acids are evolutionarily derived versus those that are ancestral. These problems can be partially resolved by using a phylogenetically-informed "vertical" approach (Fig. 1b), in which probabilistic models of sequence evolution are used to reconstruct ancestral protein sequences before and after the phylogenetic branch on which the functional change of interest occurred [1][2][3][4]. Ancestral sequences are then compared in order to reveal a set of amino acid substitutions that are phylogenetically correlated with the historic protein functional change. A major advantage of this vertical approach is that it significantly reduces the number of candidate amino acid changes relative to the horizontal approach. Despite this advantage, however, the number of candidate changes can still be sufficiently large such that it is not feasible for a molecular experimentalist to test the individual effects of every mutation in the set, let alone all combinations of historic substitutions. This problem becomes especially acute for protein functional changes that occurred on long phylogenetic branches.

Here we significantly increase the power of vertical sequence comparisons by providing an accurate means to sort historic protein sequence mutations by a set of statistical signals that, when combined, accurately discriminates known function-shifting sequence substitutions from other mutations. We describe a new method to quantify the extent to which amino acid changes between ancestors are likely to have affected protein function. We validated the accuracy of this method by using it to identify known function-shifting mutations in a training set consisting of four well-studied episodes of historic functional evolution. We show that this method accurately identified all the known functional mutations in the training set, while maintaining a low rate of false positive identification. Overall, our results strongly suggest that this method is useful for identifying the genetic basis of functional evolution in many diverse protein families.

## Results

*A Metric To Rank Ancestral Substitutions*

We developed a new metric, called $\Delta$f, to rank the mutations between any two reconstructed ancestral protein sequences, according to a composite of three statistical signatures that correlate with functional change. The metric operates on two reconstructed ancestral sequences, represented as two sequences of amino acid probability distributions (Fig. 1c). The two ancestors are separated by a phylogenetic branch on which their protein function putatively changed. The $\Delta$f metric scores the differences between the probability distributions at all homologous sequence sites in the two ancestors; sites are then ranked according to their $\Delta$f score (see Methods). The $\Delta$f scores are calculated as a composite of three signals: Kullback-Leibler divergence (k), probabilistic model violation (v), and entropic shift (d).

$$\Delta f = k \cdot v \cdot d$$

Each of the three signals in $\Delta$f measures a different aspect of functional change. The first signal, Kullback-Leibler (KL) divergence, is used in information theory to measure the information gained, in bits, between two discrete probability distributions [5]. We hypothesized that function-shifting mutations will leave a signature of higher gained information in the derived ancestor, compared to sites lacking function-shifting mutations. The second signal, probabilistic model violation, measures the extent to which the observed difference between two probability distributions deviates from our model-based expectation. Critical sites for establishing protein function are likely to experience unique biophysical constraints, and these sites may therefore exhibit mutational patterns that deviate from the patterns one would expect based on maximum likelihood phylogenetic models whose parameters have been computed from large numbers of protein families. We hypothesized that a high degree of deviation from model-based expectations could be a signature of functional evolution. The third signal, entropic shift, adjusts the sign of $\Delta$f to be negative if the mutation at a site resulted in an amino acid identity that was not conserved in descendants. This adjustment separates degenerative mutations, which may play a role in deteriorating protein function, from those mutations whose amino acid identity remained important for protein function.

Calculating $\Delta$f at every sequence site between two ancestors produces a distribution of $\Delta$f scores, and interesting mutation hypotheses are located in both tails of the distribution. Sites in the negative tail ($\Delta f < 0$) correspond to mutations that may have degenerated ancestral functions, while sites in the positive tail ($\Delta f > 0$) correspond to mutations that may have played a role in maintaining new ancestral functions. Sites with $\Delta$f values approaching zero are expected to have either uninteresting mutations, or no change at all. Examples of specific mutations and their Δf scores are shown in Fig. 1c. For instance, a mutation from a non-polar residue to a polar residue (a typical function-shifting mutation) yields high values for both KL divergence and model violation, and thus receives a high overall Δf score. In contrast, a mutation between similar distributions of aliphatic amino acids (leucine, isoleucine, and valine) yields a relatively low Δf score because there is low KL divergence between the two ancestors. In general, ancestral sites with strong support for the same amino acid before and after the phylogenetic branch of interest will yield vanishingly small Δf scores because the KL divergence is near zero and the model violation is small, while mutations between strongly-supported different amino acids will produce the highest Δf scores.

### *Identification of Known Functional Loci*

We next determined to what extent the Δf metric yields high scores for amino acid mutations that are known to have functional effects. We used a training set of four historic episodes of functional evolution in three protein families with diverse evolutionary histories in animals and fungi. For all four episodes, previous molecular experimentation identified amino acid mutations that were sufficient to have shifted protein functions from their ancestral to their derived states. We downloaded the protein sequences from these studies, reconstructed their phylogenies and ancestral sequence distributions, and then used the Δf metric to score all homologous sites between relevant ancestors (see Materials and Methods). Overall, we observed that the tails of these Δf distributions were enriched with the known functional sites, to an extent that varied among each protein family.

First, we analyzed an episode of historic evolution in Mcm1, a family of MADS-box domain transcription regulators in fungi. Mcm1 binds DNA cooperatively with other proteins in order to control the expression of genes involved in many cellular functions, including mating and arginine metabolism [6]. In the lineage leading to budding yeast *Saccharomyces cerevisiae*, there occurred a tandem duplication of Mcm1 resulting in the new paralog Arg80 (Fig. 2a, Supp. Fig. 1). After this duplication, Arg80 acquired twenty-six amino acid mutations, ten of which strongly reduced its ability to cooperatively interact with the co-factor protein MatAlpha1 and which also reduced its affinity for binding DNA [7]. These mutations are sufficient to give Arg80 its specialized properties compared with the preduplicated gene. We reconstructed the ancestral protein sequences before and after the duplication, and then used the Δf metric to score all homologous sites between the pre-duplicate ancestor (i.e. Anc.MADS) and the most-recent shared ancestor of post-duplicate Arg80 (i.e. Anc.Arg80). Compared to other mutations on this branch, the ten substitutions known to have shifted function were located in the tails of the $\Delta$f distribution. Specifically, all ten changes received $\Delta$f scores deviating from the mean by at least +/- 0.6 standard deviations; eight of the changes deviated by +/- 1.6 std. dev. (Fig. 3a). The highest-scoring mutation (glutamine to asparagine at site 42) is located on the protein surface that faces co-factor MatAlpha1 in the mating complex, and previous experiments showed that this mutation contributes to the inhibition of Arg80 to strongly bind MatAlpha1 [7]. The second and third highest scoring mutations (lysine to arginine at site 27, and glutamic acid to proline at site 7) are located on the DNA-facing side of Arg80, and these mutations were previously shown to reduce Arg80's binding affinity for DNA [7]. Overall, the $\Delta$f metric yielded relatively high scores for mutations with known effect in Mcm1, and it highly-scored mutations known to affect Arg80's DNA-binding and co-factor binding functions. These data suggest that $\Delta$f is useful for computationally recapitulating historic genetic mechanisms in the Mcm1 protein family.

We next analyzed an episode of functional evolution in the family of glucocorticoid receptors (GR), a member of a larger protein family of steroid hormone receptors (Fig. 2b, Supp. 2). In chordate animals, GR binds a wide range of glucocorticoids with varying specificity in order to regulate many cellular functions, including immune response, development, and metabolism [8]. In the evolutionary lineage leading to tetrapod and teleost species, the GR ligand-binding domain acquired thirty-seven amino acid mutations, twelve of which played critical roles in shifting the ligand-binding preference from aldosterone to cortisol. Specifically, two of the thirty-seven substitutions were necessary and sufficient to confer this shift, and another ten mutations were responsible for stabilizing the new structural conformation [9]. We calculated $\Delta$f scores for all mutations on this branch, by comparing the ancestor of all GRs (Anc.GR1) to the ancestor of tetrapod and teleost GRs (Anc.GR2). The twelve functional mutations received $\Delta$f scores deviating from the mean by at least +/- 0.6 standard deviations (Fig. 3b). Notably, the two necessary and sufficient mutations (serine to proline at site 107, and leusine to glutamine at site 111) received the highest $\Delta$f scores.

Finally, we analyzed two historic functional shifts that occurred in a family of proteins that assemble into a proton pump, the vacuolar (V-type) ATPase. In animals and Fungi, V-ATPase is an essential protein complex that acidifies many different intercellular organelles. Located in the membrane-bound domain of V-ATPase, there exists a hexameric "rotor" assembled from two proteins named VMA16 and VMA3. In the evolutionary lineage leading to Fungi, VMA3 duplicated to create a third component named VMA11 (Fig. 2c, Supp. Fig. 3). On the phylogenetic branches following this duplication, VMA3 and VMA11 independently acquired twenty-one and twenty-four amino acid substitutions, respectively. Previous work showed that seven of the twenty-one VMA3 mutations, and ten of the twenty-four VMA11 mutations caused the two subunits to degenerate their functional interactions with VMA16 and with each other. These mutations were complementary, and led to the dual retention of both VMA3 and VMA11 within fungal genomes [10]. We calculated $\Delta f$ scores for all mutations on the branch descending from the pre-duplication ancestor (Anc.3-11) to the ancestor of all post-duplicate VMA3s (Anc.3), and also on the branch descending from Anc.3-11 to the ancestor of all VMA11s (Anc.11). On the branch leading to Anc.3, all seven known function-shifting mutations received $\Delta f$ scores at least +/- 0.2 standard deviations from the mean. On the branch to Anc.11, the ten previously-identified mutations received $\Delta f$ scores at least +/- 0.1 from the mean. Taken together, the $\Delta f$ distributions for the Mcm1 family, steroid-hormone receptors, and V-ATPase subunits suggest that this metric is useful for identifying function-shifting amino acid mutations.

*Rates of Accuracy and Error*

We next determined to what extent $\Delta f$ distributions accurately discriminated known functional mutations from those mutations with little or no consequence for function. Specifically, we measured and compared the rates of true positive identification versus false positives for the protein families in the training set (Fig. 4). Overall, ranking sites according to their $\Delta f$ score resulted in the positive identification of all known function-shifting mutations in the training set with less than 9% false positive identification on the branch leading to Anc.3 in the V-ATPase family, less than 11% on the branch leading to Anc.GR2 in the steroid-hormone receptor family, and less than 13% on the branches to Anc.11 in V-ATPase and Anc.Arg80 in Mcm1 family. In other words, a ranking system based on $\Delta f$ scores produced one false hypothesis for eight true hypotheses in the worst case (on the branches leading to Anc.11 and Anc.Arg80). In the best case, $\Delta f$-based rankings produced one false hypothesis for eleven true hypotheses (on the branch leading to Anc.Arg80).

*Comparison to Other Ranking Methods*

The $\Delta f$ -based ranking method was more accurate at predicting functional loci than a previously-published method, DIVERGE. This previous method was designed to identify so-called type-I and type-II functional loci [11]. Type-I loci are characterized by shifts in evolutionary rates, typically due to strong amino acid conservation in species with a function of interest and poor conservation in species without the function [12]. Type-II loci are characterized by shifts in biochemical amino acid properties, such as changes to hydrophobicity and polarity [13]. We used DIVERGE to identify type-I and type-II functional loci on the branch leading to Anc.Arg80 in the Mcm1 protein family. We then compared the type-I and type-II scores to their corresponding $\Delta f$ scores. We observed that type-I scores correlated poorly with $\Delta f$ scores (R=0.11), while type-II scores correlated well with $\Delta f$ scores (R = 0.81) (Supp. Fig. 5a,b). We next compared the rates of true and false positive identification for type-I, type-II, and $\Delta f$-based rankings. While the $\Delta f$-based method identified 100% of the known true positives with only 13% false positive identification rate, DIVERGE achieved 100% true positive identification with 83% false positives for type-1, and 100% true positive rate with 62% false positives for type-II (Supp. Fig. 5c). These results suggest that an approach to identifying functional loci based on information theory (*i.e.,* the $\Delta f$-based method) is superior to identification methods based on shifts in evolutionary rates (*i.e.,* DIVERGE type-I), and methods based on ad-hoc biochemical categorization of amino acid properties (*i.e.,* DIVERGE type-II).

*Analysis of Branches Lacking Functional Evolution*

When the $\Delta f$ metric was used to rank amino acid mutations on phylogenetic branches lacking any known changes to protein function, the method produced $\Delta f$ distributions that were relatively smaller than the distributions produced on branches with known functional evolution (Supp. Fig. 6). Although no functional changes are thought to have occurred on these branches, their $\Delta f$ distributions did contain a few high-scoring sites. This indicates either that $\Delta f$ is prone to small numbers of false-positive predictions on these branches, or that these branches correlate with some degree of cryptic functional evolution.

In the Mcm1 protein family, previous functional data suggests that no major changes occurred on the phylogenetic branch descending from Anc.Arg80 to the *S. cerevisiae* clade (Supp. Fig. 1). On this branch, thirteen sequence sites experienced amino acid substitutions, but only four of these changes received $\Delta f$ scores exceeding +/- 1.0 standard deviations from the mean. In contrast, on the branch connecting Anc.MADS to Anc.Arg80 (i.e., a known function-changing branch), eleven of twenty-six mutations received $\Delta f$ scores exceeding 1.0 s.d. from the mean (Supp. Fig. 6a).

In the steroid-hormone receptor family, ligand-binding specificities of present-day GRs suggest that ligand-binding specificity was essentially conserved on the branch descending from Anc.GR2 to mammalian glucocorticoid receptors (Supp. Fig. 2). Fourteen amino acid sites mutated on this branch, but only six of those sites received $\Delta f$ scores exceeding 1.0 s.d. from the mean (Supp. Fig. 6b). In contrast, the branch connecting Anc.GR1 to Anc.GR2, on which protein function is known to have changed, contains 37 mutations of which 15 have Δf scores exceeding 1.0 s.d. from the mean.

Finally, in the V-ATPase proteolipid subunit family, there is little evidence that VMA3 function changed on the branch leading from Anc.VMA3 to the ancestor of Ascomycete yeast (Supp. Fig. 3). Thirteen amino acid substitutions occurred on this branch, but only two sites received Δf scores greater than two s.d. from the mean (Supp. Fig. 6c). In contrast, the branch connecting Anc.3-11 to Anc.3 (on which a functional change occurred) contains seven mutations with Δf scores exceeding 1.0 s.d. from the mean. Taken together, our analysis of Δf scores on phylogenetic branches putatively lacking any functional evolution suggests that the Δf metric could be useful in blind screens to find phylogenetic branches that are relatively enriched with functional changes. These data also suggest that some low degree of cryptic functional evolution may have occurred on those branches previously thought to lack functional change.

## Discussion

Identifying historical amino acid substitutions responsible for the acquisition of present-day functions of proteins is a general problem in molecular biology. By taking advantage of the additional information present in computationally reconstructed ancestral amino acid probability distributions, we developed a new method, named Δf, to score and rank amino acid protein sequence sites according to their signature of functional evolution. This approach correctly identified the known functional loci in a training set of three diverse protein families, while maintaining a low rate of false positive identification. Our results strongly suggest that the Δf metric can be a useful tool for generating and prioritizing specific hypotheses about historic functional evolution in a wide variety of protein families.

One advantage of the Δf-based ranking method is that it can discriminate between historic mutations that resulted in a gain of amino acid conservation versus a loss of conservation. By adjusting the sign of Δf based on the direction of entropic shift between two ancestors, our approach provides experimentalists two useful indicators. First, the absolute magnitude of Δf can be used to rank sites according to their historic function-shifting potential, while the sign (positive vs. negative) of Δf indicates whether the change led to a gain of conservation (indicated by $\Delta f > 0$) or loss of conservation (indicated by $\Delta f < 0$). It is unknown in what proportion functional evolution is driven by gains versus losses of conservation, and we predict that experimentalists would be interested in testing candidate mutations at both tails of a Δf distribution.

The training set used in this study has limitations. Although the historic function-shifting episodes in the training set were each isolated to a single phylogenetic branch, it is unknown if other functional shifts also occurred on that same branch. It is possible that some of the high-scoring mutations that seem to be false positives in our study may actually be true positives, responsible for shifting other, yet unknown, protein functions. Therefore, the rates of false positive identification reported in this study should be interpreted as upper bounds on the real false positive rates.

The $\Delta f$-based method does not account for functional shifts caused by insertions or deletions (i.e., indels) of sequence sites. The $\Delta f$ metric cannot be applied at indel sites because there does not exist an amino acid probability distribution for the ancient ancestor (in the case of an insertion), or for the derived ancestor (in the case of a deletion). The frequencies and functional effects of indel events are difficult to predict, but recent work has proposed a promising phylogenetic Poisson-based model of indels [14]. One future direction would be to use this type of approach to compute model-based expectations for indel events, which could then be compared to empirical observations in order to identify indel events that deviate from expectations. In the meantime, our software implementation of the $\Delta f$ metric currently reports a list of indel events that occurred on the phylogenetic branch of interest. Experimentalists are encouraged to examine these indel events alongside their high-scoring $\Delta f$-based predictions in order to consider a comprehensive set of mutation hypotheses.

The $\Delta f$-based method for ranking function-shifting mutations should not be confused with methods for identifying sites under positive selection. The $\Delta f$ method identifies historic mutations that likely contributed to functional change, while making no assumptions about the adaptive consequences of those mutations. An open question in evolutionary theory is whether functional protein evolution is primarily driven by mutations that offer a fitness advantage, versus mutations that are non-adaptive [15] [16]. This question can be addressed by identifying functional loci, using $\Delta f$ and other methods, and then testing the fitness effects of mutations at those loci.

We note that the approach described here does not require that the specific consequence of the function-shifting amino acid substitutions be known ahead of time. The method simply identifies (through extreme $\Delta f$ values along a particular lineage) amino substitutions that have a high probability of changing the function of the encoded protein. We believe that a particular strength of the approach is that it allows the consequence of a function-shifting amino acid substitution to be experimentally tested. In lineages that contain a genetically tractable organism, the amino acid substitutions identified through extreme $\Delta f$ values can be experimentally reverted to the ancestral state, and the effects can then be observed. In this way – particularly if something is known about the protein in which the historic changes occurred – the consequences of function-shifting mutations can be directly assessed.

The relationship between changes to genome sequence and changes to organismal function has been characterized in a relatively small number of species. It remains unknown to what extent our present-day understanding of mutational effect size, pleiotropy, and epistasis represent a general pattern for biology, or are unique peculiarities of model systems. Our hope is that the $\Delta$f metric will facilitate rapid discovery of function-shifting genetic loci in a wide range of protein families in both model and non-model species. The $\Delta$f metric can be combined with complementary approaches to identify functional loci that occurred over coalescent timescales [17], and functional loci that are co-evolving within coding sequences [18]. Taken together, it now seems possible to use this larger class of computational tools to automatically predict functional loci in a wide range of genomic contexts, and ultimately to acquire a more comprehensive understanding of functional evolution across the tree of life.

## Methods and Materials

*Sequence Collection and Phylogenetic Analysis*

Amino acid sequences for the three protein families studied in this paper (Mcm1, steroid-hormone receptors, and V-ATPase subunits) were acquired from collections curated in their previously-published analyses ([7], [9], [10]). The sequences in each family were aligned using three different software methods: MUSCLE [19], MSAProbs [20], and PRANK [21]. For each alignment, twelve different likelihood-based evolutionary models were tested for their goodness-of-fit using the Akaike information Criterion [22]. The twelve models included all combinations of three options: the amino acid substitution matrix (JTT [23], WAG [24], or LG [25]), the use of a multiple gamma-distributed evolutionary rates (yes/no), and the use of amino acid stationary frequencies that are either fixed or estimated. Statistical support for branches was computed as approximate likelihood ratios (aLRs), which express the likelihood (L1) of the branch divided by the likelihood (L2) of the next-best tree lacking that branch. aLR values were computed by using PhyML [26], [27] to first compute the aLR test statistic (aLRT), and then coverting the aLRT to aLR as follows.

$$aLR = \frac{L1}{L2} = e^{\left(\frac{aLRT}{2}\right)}$$

*Ancestral Sequence Reconstruction*

Ancestral amino acid probability distributions were reconstructed for all sequence sites at all internal phylogenetic nodes for the three protein families studied in this paper, using the software package Lazarus to control PAML [4], [28]. For each protein family, thirty-six different reconstructions were performed, corresponding to all combinations of the three sequence alignments and the twelve evolutionary models described above. Insertion/deletion characters were added using Fitch's parsimony [29], based on the following pre-defined "outgroup" sequences: the *Y. lipolytica* sequence for the Mcm1 family, AR and PR sequences for the steroid-hormone receptor family, and VMA16 sequences for the V-ATPase subunit family.

*Identification and Ranking of Functional Amino Acid Substitutions*

In order to identify and rank functional loci, homologous sequence sites were compared between reconstructed ancestors before and after each of the four functional shifts studied in this paper (Fig. 1). The comparison is described as follows for a reconstructed ancestral sequence *x*, and its descendant ancestral sequence *y*. Both sequences contain *N* amino acid sites. In both ancestors, every site, *i,* is represented as a probability distribution with twenty values, one for every possible amino acid. For example, the probability $P(x_i=a)$ is the probability that site *i* in ancestor *x* was amino acid *a*. At each site that is homologous between *x* and *y*, probability distributions were compared and scored according to the $\Delta f$ metric, which is the multiplicative product of three measures: Kullback-Leibler divergence (*k*), probabilistic model violatation (*v*), and entropic shift (*d*).

$$\Delta f(x_i, y_i) = k(x_i, y_i) \cdot v(x_i, y_i) \cdot d(x_i, y_i)$$

*Kullback-Leibler Divergence*

Kullback-Leibler (KL) divergence is a measure of the information gain in one discrete probability distribution, given another probability distribution [5]. The KL divergence between amino acid probability distributions was calculated as follows. For two distributions, *x* and *y,* at a single site *i,* the KL divergence between the distributions is expressed as the function $k(xi, yi)$, which is computed as a sum for each amino acid, *a,* in the set, *A,* of all possible amino acids. For each *a,* the partial KL sum is the product of the posterior probability that ancestor *x* was amino acid *a* at site *i,* times the log of the fraction of the probability that *x* was *a* at *i*, divided by the probability that *y* was *a* at site *i.*

$$k(x_i, y_i) = \sum_{a}^{\in a \subset A} P(x_i = a) \times \log\left(\frac{P(x_i = a)}{P(y_i = a)}\right)$$

*Probabilistic Model Violation*

The extent to which the observed differences between two ancestors matched model-based expectations was computed as follows. For two reconstructed ancestors, *x* and *y,* the extent of their observed model violation at a single site, *i,* was computed using the function *v,* assuming a phylogenetic branch length *t* that separates ancestors *x* and *y*, and a Markovian substitution matrix *M*, expressing the relative substitution rates between amino acids. For example, *M* could be a well-known matrix such as JTT or WAG. The extent of model violation was computed as a sum for all 380 (i.e., 20x19) possible mutations between amino acids *a* in *x* to *b* in *y*. The partial sum for each possible mutation was calculated as the square difference between the expected and observed mutational distances of that the probabilities underlying that mutation. The expected mutational distance is computed from the Poisson process: the probability of amino acid *a* existing at site *i* in ancestor *x,* multiplied by the exponent of the branch length *t* times the relative substitution rate, *M[a,b],* between amino acids *a* and *b.* The observed mutational distance is the probability of state *a* existing at site *i* in *x,* multiplied by the probability of mutated state *b* existing at site *i* in *b.*

$$v(x_i, y_i | t, M) = \sum_{a}^{\in a \subset A} \sum_{b}^{\in b \neq a, b \subset A} \left(\left[p(x_i = a) \times e^{tM[a,b]}\right] - \left[p(x_i = a) \times p(y_i = b)\right]\right)^2$$

*Entropic Shift*

In order to differentiate amino acid mutations that led to a gain of conservation versus those that led to loss of conservation, the shift in entropy between two amino acid probability distributions was calculated as follows. The entropy of an amino acid probability distribution, for an ancestor *x* at site *i,* is computed as a sum for each amino acid, *a,* equal to the probability of amino acid *a* existing at site *i* in *x*, times the log of the probability of *a* existing at *i* in *x*. The sign, *d*, is made equal to 1.0 if the derived ancestor, *y*, decreased in entropy. Conversely, *d* is set to minus 1 if the derived ancestor increased in entropy.

$$entropy(x_i) = \sum_{a}^{\in a \subset A} p(x_i = a) \times \log\big(p(x_i = a)\big)$$

$$d(x_i, y_i) = \begin{cases} 1, if\ entropy(x_i) \geq entropy(y_i) \\ -1, if\ entropy(x_i) < entropy(y_i) \end{cases}$$

*Bayesian Incorporation of Alignment and Phylogenetic Uncertainty*

In order to reduce the effects of error and uncertainty in the underlying sequence alignments and phylogenetic trees, we averaged $\Delta f$ scores from ancestors inferred by the thirty-six unique combinations of sequence alignment methods and phylogenetic models for each protein family in our analysis (Supplemental Fig. 6). We weighted the $\Delta f$ scores from each alignment by the posterior probability of their maximum likelihood model, relative to other models from the same alignment. We then weighted the three alignment methods equally.

*Visualization of Predicted Loci*

The $\Delta f$ scores for each functional episode were plotted across ancestral sequence sites. The scores for each episode were also plotted as a distribution by binning the mutations according to the $\Delta f$ score, and then counting the number of mutations in each bin.

*Measuring Accuracy and Error*

The rates of true positive identification and false positive identification were measured and compared for the $\Delta f$ distributions for each historic function shift in the training set. For each distribution, the $\Delta f$ scores were sorted in descending order by their absolute value. A set of putative mutations was then incrementally built, starting with the first $\Delta f$ score and proceeding in order. At each increment, the true positive rate was measured as the proportion of known functional mutations in the putative set, relative to the total number of known functional mutations. Similarly, the false positive rate was measured as the proportion of non-functional mutations in the putative set, relative to the total number of non-functional mutations. Sites with no change in their maximum likelihood amino acid between the pre- and post-shift ancestor were excluded from the analysis.

*Comparison to DIVERGE*

For the Mcm1 protein family, we compared the $\Delta f$ scores to the scored calculated by DIVERGE for type-I and type-II functional evolution. We used the default settings in DIVERGE, with the maximum likelihood Mcm1 phylogeny generated from the MSAProbs alignment algorithm.

## Acknowledgements

The data reported in this paper are tabulated in Supplemental Materials. This study was supported by National Institute of Health Grant XX and XX. We thank Trevor Sorrells and Katie Pollard for useful discussions about this project. We thank members of the Johnson lab at UCSF for helpful comments on the manuscript.

## References

1. Yang, Z., Kumar, S., & Nei, M. (1995) A New Method of Inference of Ancestral Nucleotide and Amino Acid Sequences. *Genetics* 141, 1641--1650.
2. Thornton, J. W. (2004) Resurrecting Ancient Genes: Experimental Analysis of Extinct Molecules. *Nature* 5, 366-375.
3. Liberles, D. (editor) (2007) Ancestral Sequence Reconstruction. Oxford University Press.
4. Hanson-Smith, V., Kolaczkowski, B., & Thornton, J. W. (2010) Robustness of Ancestral Sequence Reconstruction to Phylogenetic Uncertainty. *Molecular Biology and Evolution* 27
5. Kullback, S. (1959) Information theory and statistics. Dover Publications.
6. Messenguy, F. & Dubois, E. (2003) Role of MADS box proteins and their cofactors in combinatorial control of gene expression and cell development. *Gene* 316, 1-21.
7. Baker, C. R., Hanson-Smith, V., & Johnson, A. D. (2013) Following gene duplication, paralog interference constrains transcriptional circuit evolution. *Science* 342, 104-8.
8. Carroll, S., Ortlund, E. A., & Thornton, J. W. (2011) Mechanisms for the Evolution of a Derived Function in the Ancestral Glucocorticoid Receptor. *PLoS Genetics* 7(6).
9. Bridgham, J. T., Ortlund, E. A., & Thornton, J. W. (2009) An epistatic ratchet constrains the direction of glucocorticoid receptor evolution. *Nature* 461, 515-9.
10. Finnigan, G. C., Hanson-Smith, V., Stevens, T. H., & Thornton, J. W. (2012) Evolution of increased complexity in a molecular machine. *Nature* 481, 360-4.
11. Gu, X., Zou, Y., Su, Z., Huang, W., Zhou, Z., Arendsee, Z., & Zeng, Y. (2013) An Update of DIVERGE Software for Functional Divergence Analysis of Protein Family. *Mol Biol Evol* , .
12. Gu, X. (1999) Statistical methods for testing functional divergence after gene duplication. *Mol Biol Evol* 16, 1664-74.
13. Gu, X. (2006) A simple statistical method for estimating type-II (cluster-specific) functional divergence of protein sequences. *Mol Biol Evol* 23, 1937-45.
14. Bouchard-Côté, A. & Jordan, M. I. (2012) Evolutionary inference via the Poisson Indel Process. *Proc Natl Acad Sci U S A* , .
15. Bloom, J. D. & Arnold, F. H. (2009) In the light of directed evolution: pathways of adaptive protein evolution. *Proc Natl Acad Sci U S A* 106 Suppl 1, 9995-10000.
16. Barrett, R. D. H. & Hoekstra, H. E. (2011) Molecular spandrels: tests of adaptation at the genetic level. *Nat Rev Genet* 12, 767-80.
17. Grossman, S. R., Shlyakhter, I., Shylakhter, I., Karlsson, E. K., Byrne, E. H., Morales, S., Frieden, G., Hostetter, E., Angelino, E., Garber, M., Zuk, O., Lander, E. S., Schaffner, S. F.,

& Sabeti, P. C. (2010) A composite of multiple signals distinguishes causal variants in regions of positive selection. *Science* 327, 883-6.

18. Lockless, S. W. & Ranganathan, R. (1999) Evolutionarily conserved pathways of energetic connectivity in protein families. *Science* 286, 295-9.
19. Edgar, R. C. (2004) MUSCLE: multiple sequence alignment with high accuracy and high throughput. *Nucleic Acids Research* 32, 1792--1797.
20. Liu, Y., Schmidt, B., & Maskell, D. L. (2010) MSAProbs: multiple sequence alignment based on pair hidden Markov models and partition function posterior probabilities. *Bioinformatics* 26, 1958-64.
21. Loytynoja, A. & Goldman, N. (2008) Phylogeny-Aware Gap Placement Prevents Errors in Sequence Alignment and Evolutionary Analysis. *Science* 320, 1632--1635.
22. Akaike, H. (1973) Information Theory and an Extension of the Maximum Likelihood Principle. *Proceedings of the 2nd International Symposium on Information Theory* , 267--281.
23. Jones, D. T., Taylor, W. R., & Thornton, J. M. (1991) The rapid generation of mutation data matrices from protein sequences. *Bioinformatics* 8, 275-282.
24. Whelan, S. & Goldman, N. (2001) A general empirical model of protein evolution derived from multiple protein families using a maximum-likelihood approach. *Molecular Biology and Evolution* 18, 691-9.
25. Le, S. Q. & Gascuel, O. (2008) An improved general amino acid replacement matrix. *Molecular Biology and Evolution* 25, 1307-20.
26. Guindon, S., Dufayard, J. F., Lefort, V., Anisimova, M., Hordijk, W., & Gascuel, O. (2010) New Algorithms and Methods to Estimate Maximum-Likelihood Phylogenies: Assessing the Performance of PhyML 3.0. *Systematic Biology* 59, 307--321.
27. Guindon, S., Dufayard, J. F., Lefort, V., Anisimova, M., Hordijk, W., & Gascuel, O. (2010) New algorithms and methods to estimate maximum-likelihood phylogenies: assessing the performance of PhyML 3.0. *Syst Biol* 59, 307-21.
28. Yang, Z. (2007) PAML 4: Phylogenetic Analysis by Maximum Likelihood. *Molecular Biology and Evolution* 24, 1586--1591.
29. Fitch, W. M. (1971) Toward Defining the Course of Evolution: Minimum Change for a Specific Tree Topology. *Systematic Zoology* 20, 406--416.

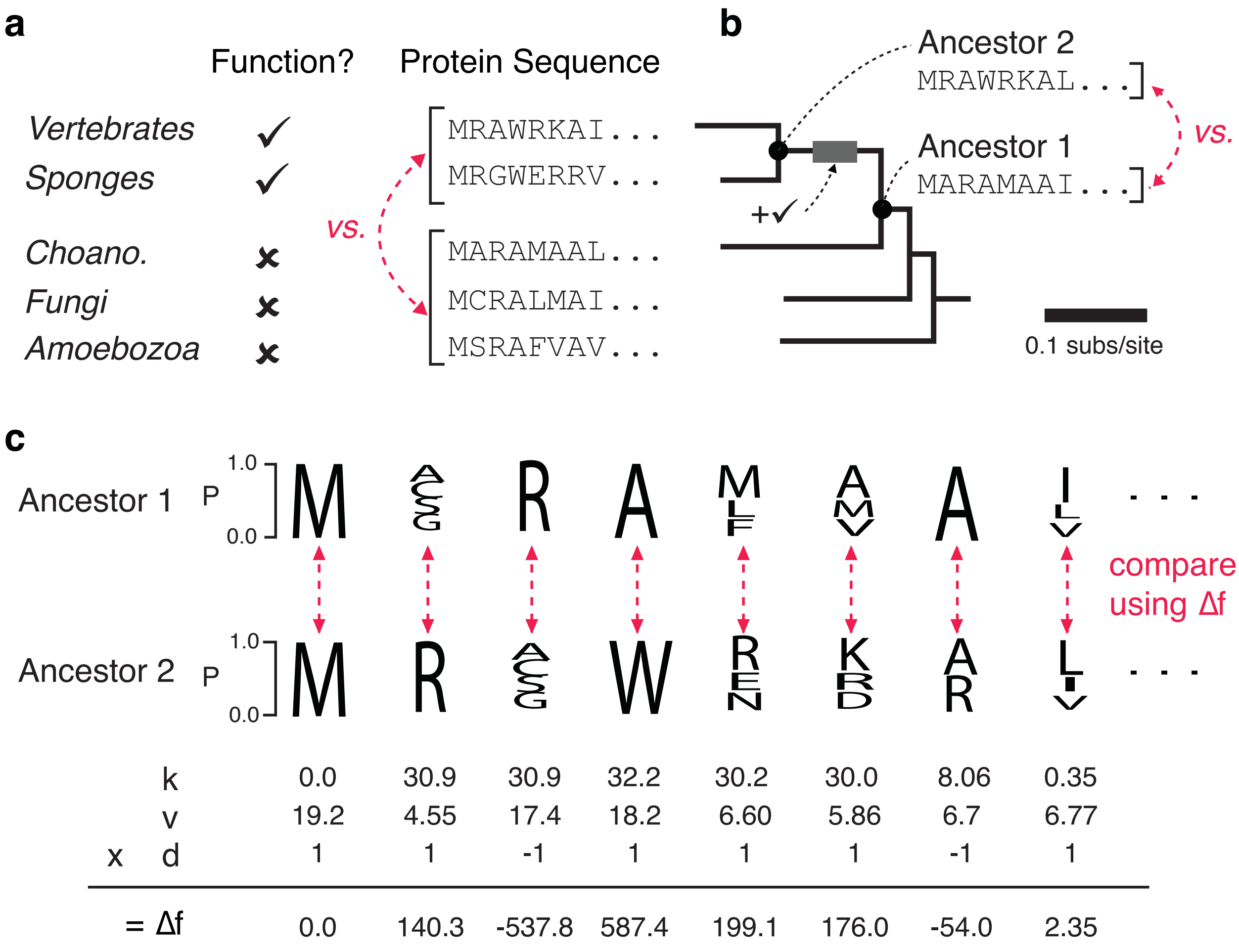

**Figure 1. Examples of vertical and horizontal comparison approaches to identify functional amino acid differences.** In this hypothetical example, vertebrates and sponges possess a particular protein function, while Choanoflagellates, Fungi, and Amoebozoa lack the function. (a) A horizontal approach directly compares protein sequences between species with different functions. (b) A vertical approach isolates the functional change to a single phylogenetic branch, reconstructs ancestors on either side of that branch, and then compares their reconstructed sequences. (c) The Δf metric extends the vertical approach, and compares amino acid probability distributions between reconstructed ancestors in order to rank their mutations. In the distributions shown, the height of the amino acid corresponds to the posterior probability of that residue existing at the sequence position in the ancestor. Underneath the distributions, scores are shown for KL divergence *k*, probabilistic model violation *v*, and entropic shift *d*. The Δf score computed as the product of *k, v,* and *d.*

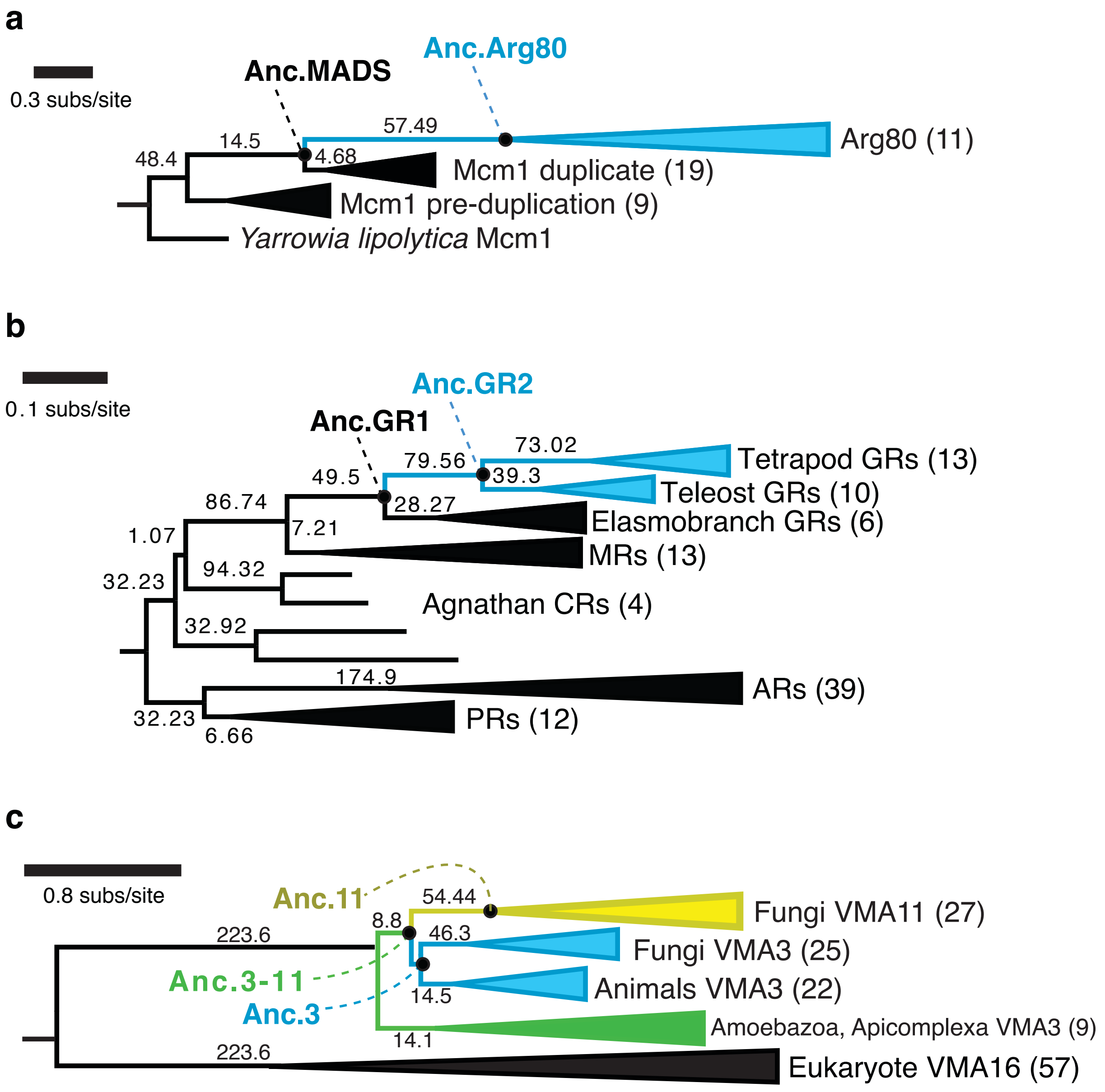

**Figure 2. Maximum likelihood phylogenies for three protein families with histories of functional evolution.** (a) In the Mcm1 family, the ancestral protein Anc.MADS duplicated into Arg80 and the Mcm1 duplicate. The most-recent shared ancestor of all Arg80 paralogs is labeled as Anc.Arg80. The tree was built from MADS-box domain sequences. (b) In the steroid hormone receptor family, the ligand-binding preference changed along the branch connecting two ancestral glucocorticoid receptors, named Anc.GR1 and Anc.GR2. (c) In the family of V-ATPase proteolipid subunits, functional evolution occurred on the branches descending from the ancestral protein Anc.3-11 to its duplicated descendants Anc.11 and Anc.3. All three phylogenies were inferred from amino acid sequence alignments. Branch lengths express substitutions per sequence site. Branch support values express approximate likelihood ratios between the phylogenetic branch and the next-best phylogenetic hypothesis lacking that branch. Values in parenthesis express the number of sequences sampled in each phylogenetic clade. Colors indicate clades with derived protein functions.

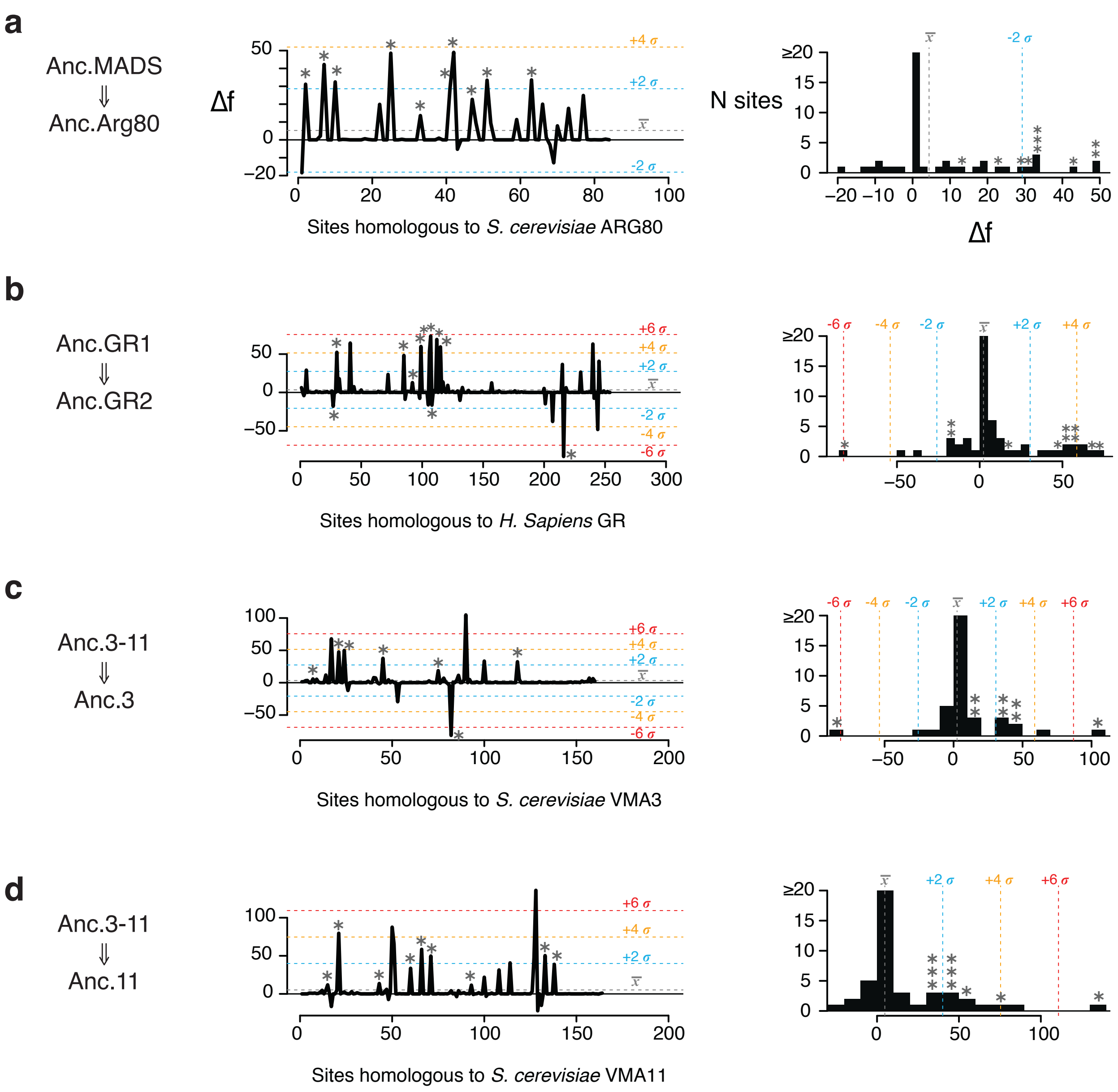

**Figure 3. Predicted loci of functional evolution in four protein lineages.** (a) The $\Delta f$ scores on the branch leading from Anc.MADS to Anc.Arg80 in the Mcm1 protein family are plotted for every amino acid site homologous to the *S. cerevisiae* ARG80 sequence. The $\Delta$f peaks for ten experimentally verified mutations are marked with asterisks (*). The histogram, on the right side, illustrates the same data binned into groups by $\Delta$f score, with bars expressing the number of sites counted in each bin. The $\Delta$f scores are shown similarly for mutations in other protein families on the phylogenetic branches leading from (b) the ligand-binding domains of Anc.GR1 to Anc.GR2 in the steroid hormone receptor family, (c) Anc.3.11 to Anc.3 in the V-ATPase proteolipid subunit family, and (d) Anc.3-11 to Anc.11. Bars in the histograms are colored according to their divergence from the mean $\Delta f$ score. Blue is +/-2 standard deviation, orange is +/-4 s.d., and red is +/- 6 s.d.

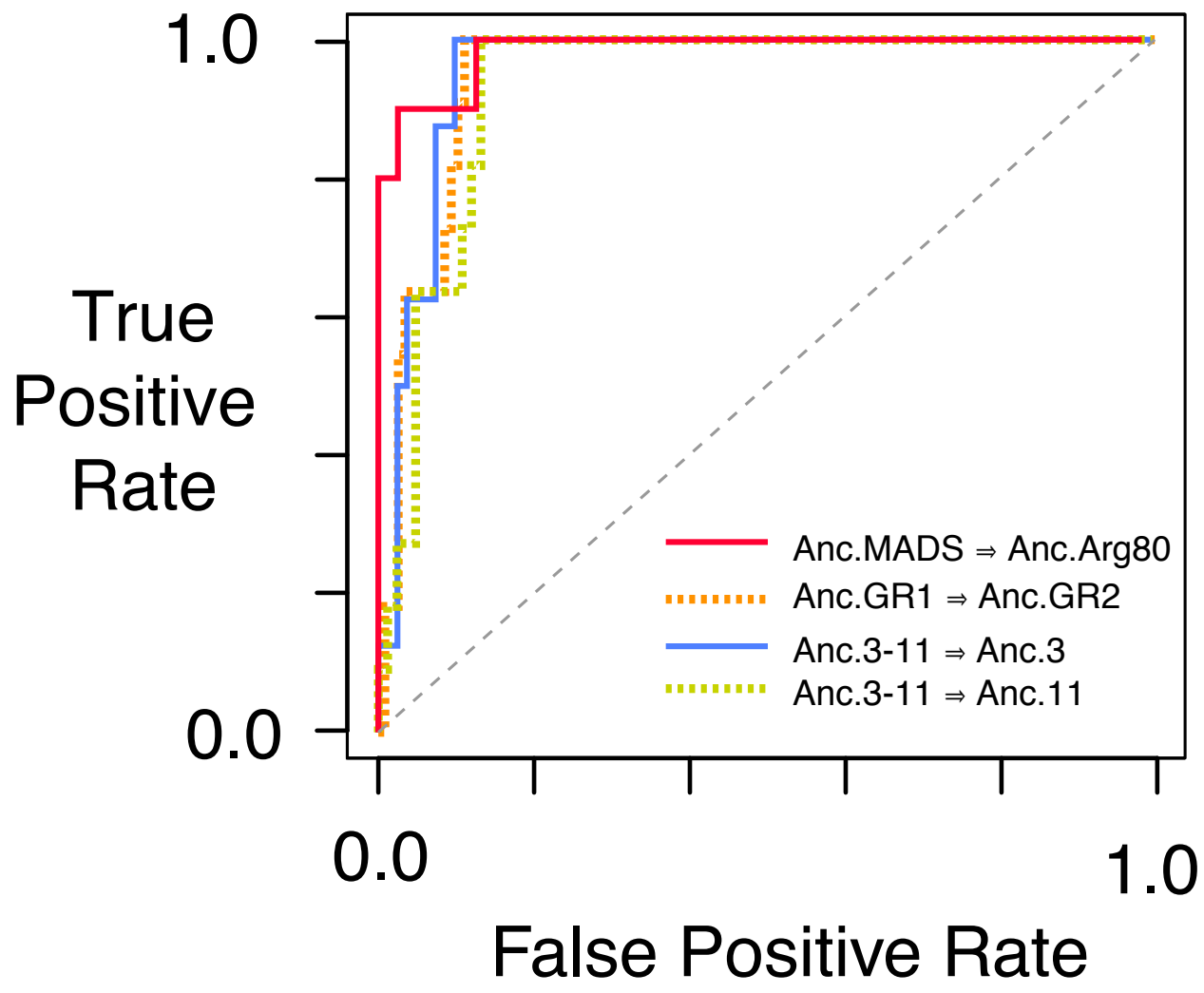

**Figure 4. Accuracy and error of Δf-based predictions.** The false positive rate is plotted against the true positive rate for predicted loci of evolution in four historic functional shifts. Each series is colored to express data from different protein lineages. The diagonal line expresses the expected ratios of true/false rates for a method with no predictive power.

# Supplemental Materials

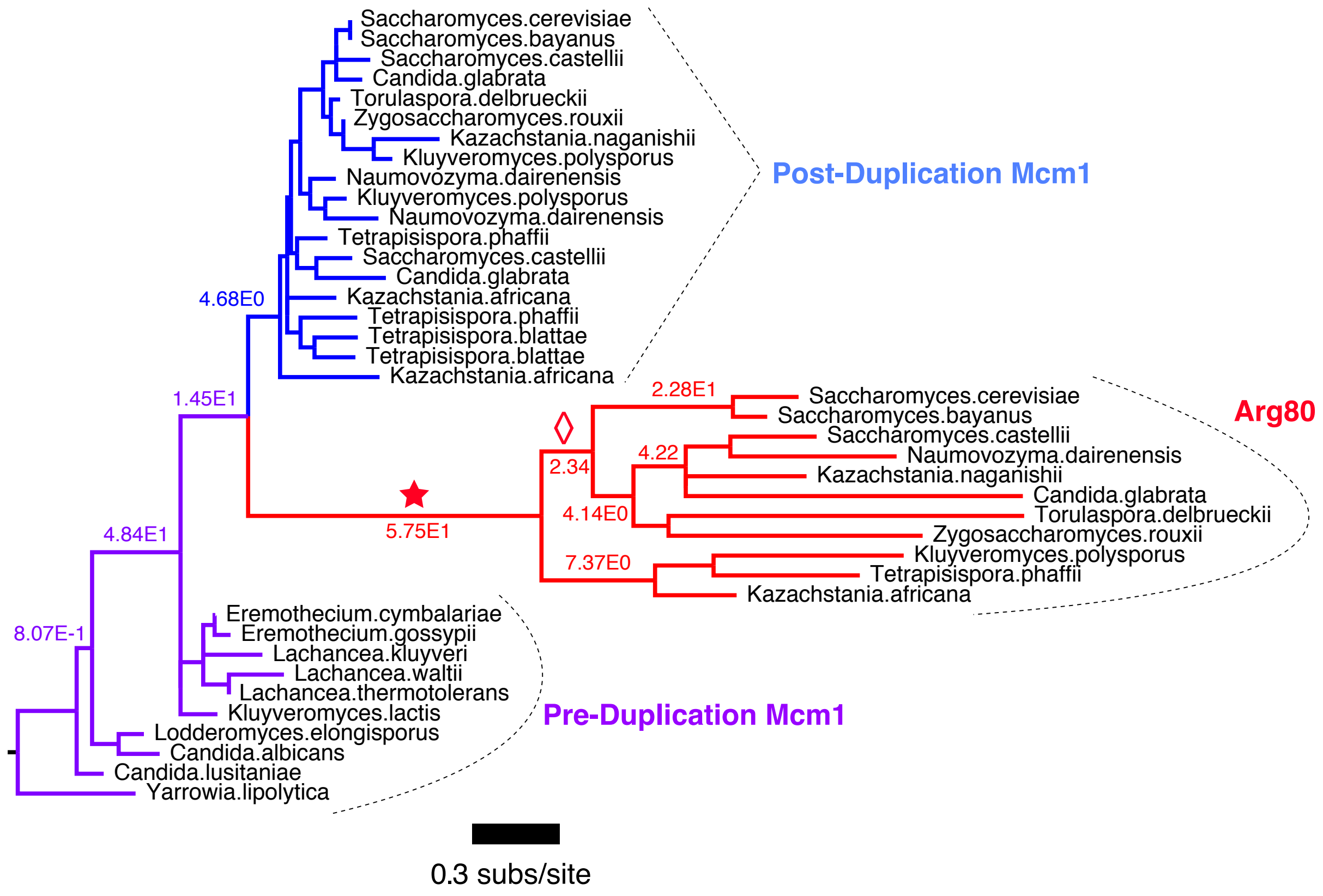

**Supplemental Figure 1. Maximum likelihood phylogeny of Mcm1 sequences in Ascomycete yeast.** Three major groups of sequences are indicated: Pre-Duplication Mcm1, Post-Duplication Mcm1, and Arg80. Branch lengths express substitutions per sequence site. Branch support values express approximate likelihood ratios between the shown phylogenetic branch and the next-best hypothesis lacking that branch. The star indicates the branch with the functional change studied in this paper. The diamond indicates a control branch also studied.

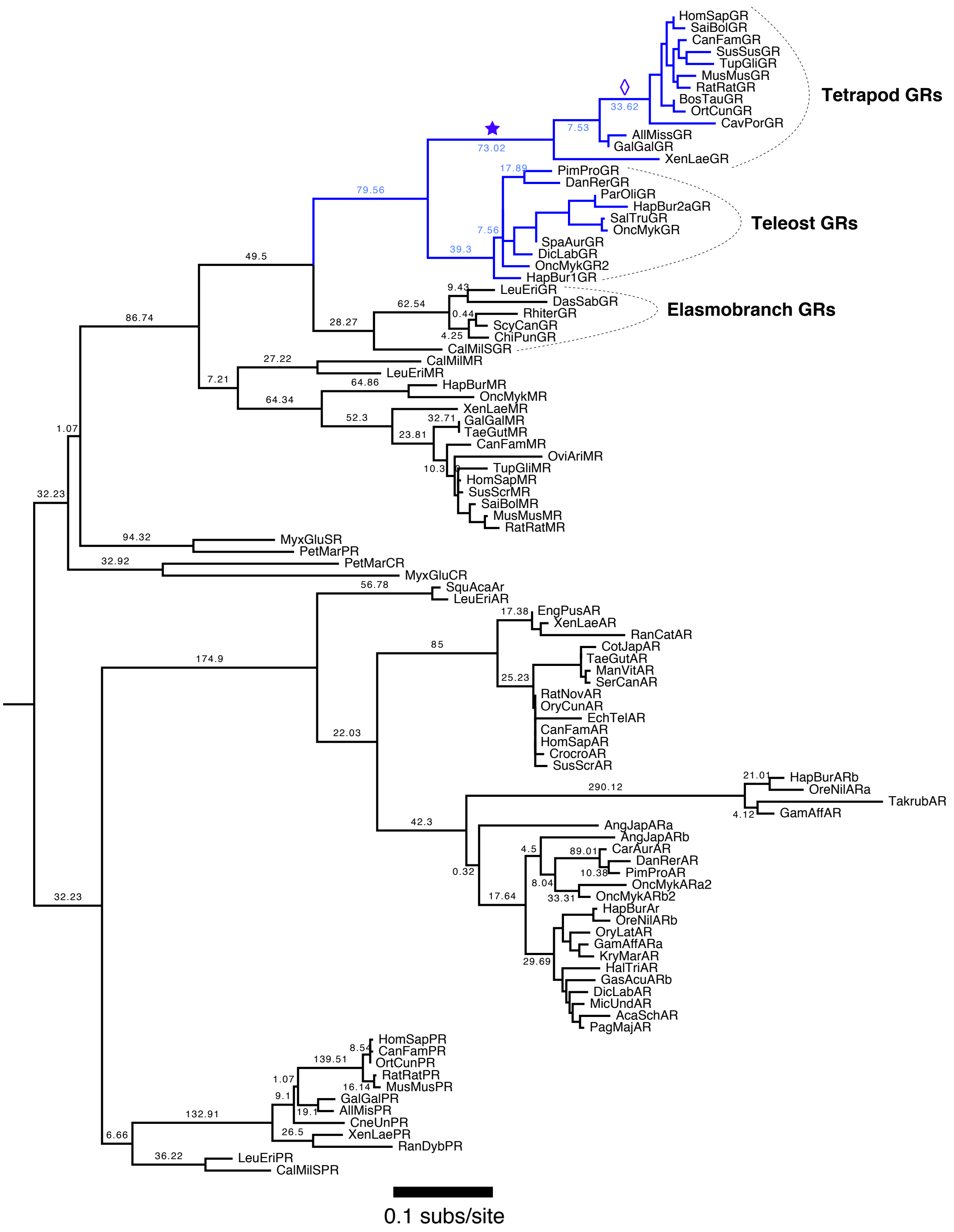

**Supplement Figure 2. Maximum likelihood phylogeny of steroid hormone receptor ligand-binding domain sequences.** Three major groups of glucocorticoid (GR) sequences are indicated: Elasmobranch GRs, Teleost GRs, and Tetrapod GRs. Branch lengths express substitutions per sequence site. Branch support values express approximate likelihood ratios between the shown phylogenetic branch and the next-best hypothesis lacking that branch. The star indicates the branch with the functional change studied in this paper. The diamond indicates a control branch also studied.

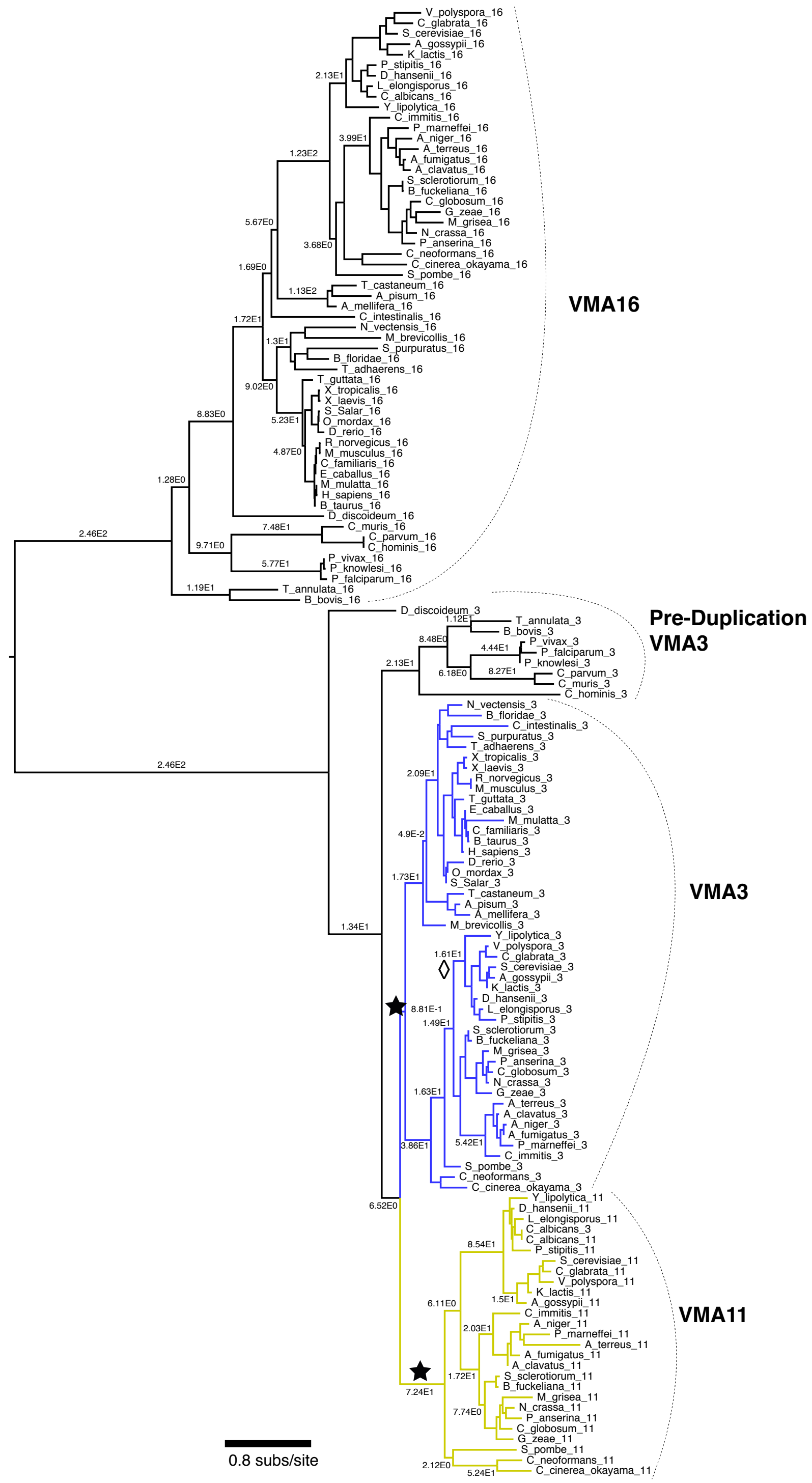

**Supplemental Figure 3. Maximum likelihood phylogeny of V-ATPase subunits VMA16, VMA3, and VMA11 in Opisthokonts.** Branch lengths express substitutions per sequence site. Branch support values express approximate likelihood ratios between the shown phylogenetic branch and the next-best hypothesis lacking that branch. The two stars indicates the branches with functional changes studied in this paper. The diamond indicates a control branch also studied.

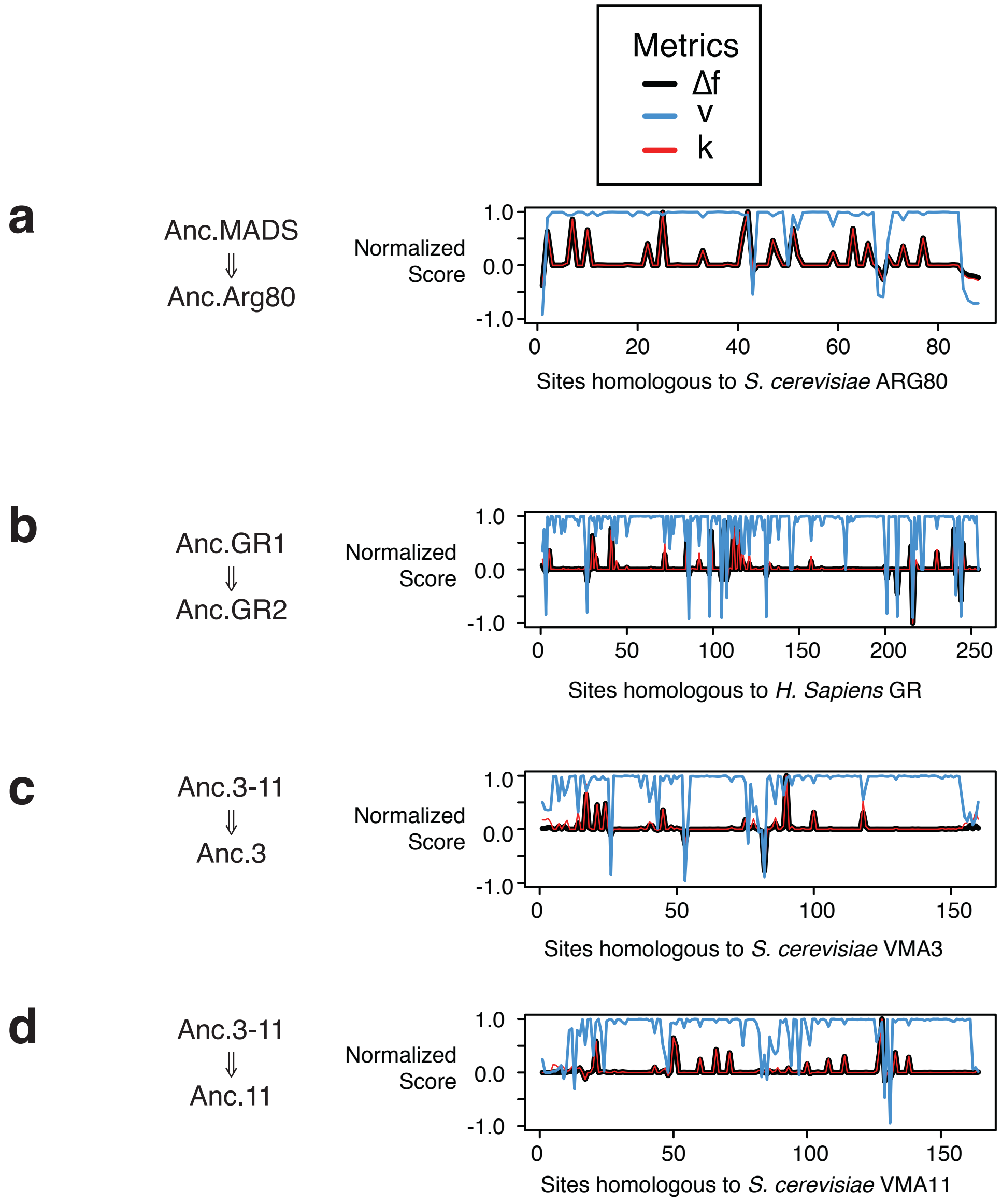

**Supplement Figure 4. Composite site scores in four protein lineages.**

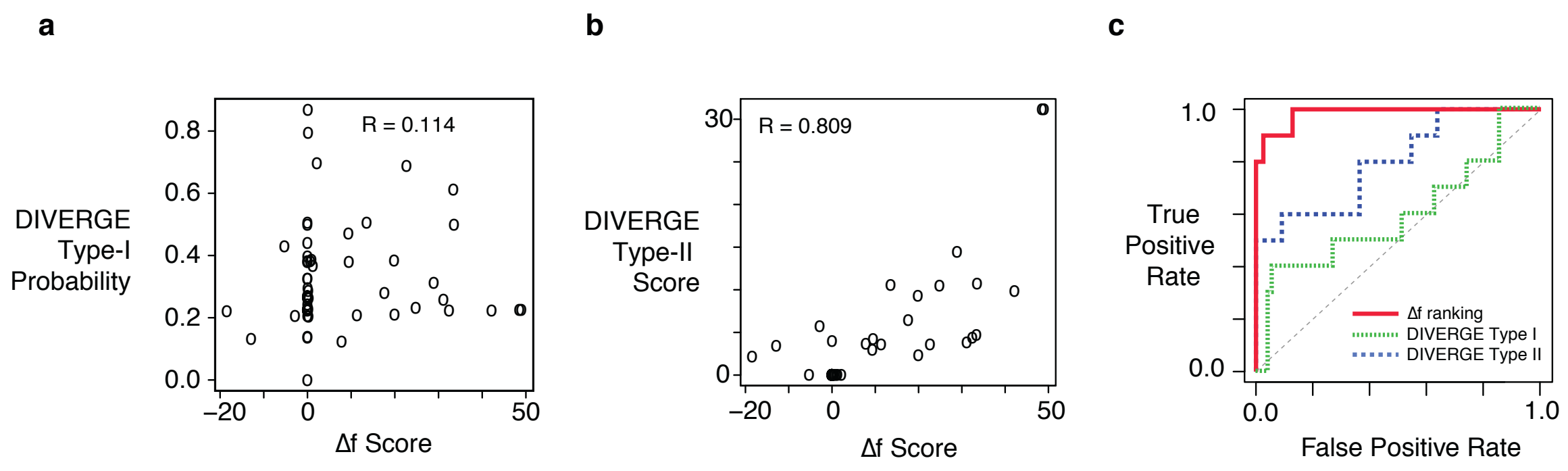

**Supplemental Figure 5. Comparison of Δf scores to DIVERGE scores.** Data is shown for amino acid mutations on the branch leading to Anc.Arg80 in the Mcm1 protein family. (a) The Δ*f* score for every mutation on the branch leading to Anc.Arg80 is compared to the probability that the corresponding sequence site is a type-I functional loci, according to DIVERGE. ***R*** is the Spearman's rank coefficient. (b) Δf scores are compared to the type-II functional loci score for each site. (c) The false positive rate is plotted against the true positive rate for predicted loci of evolution, similar to Figure 4 in the main text. The false/true data for the Δf-based ranking method is shown as a red solid line, and the data for DIVERGE-based rankings are shown in green and blue dashed lines.

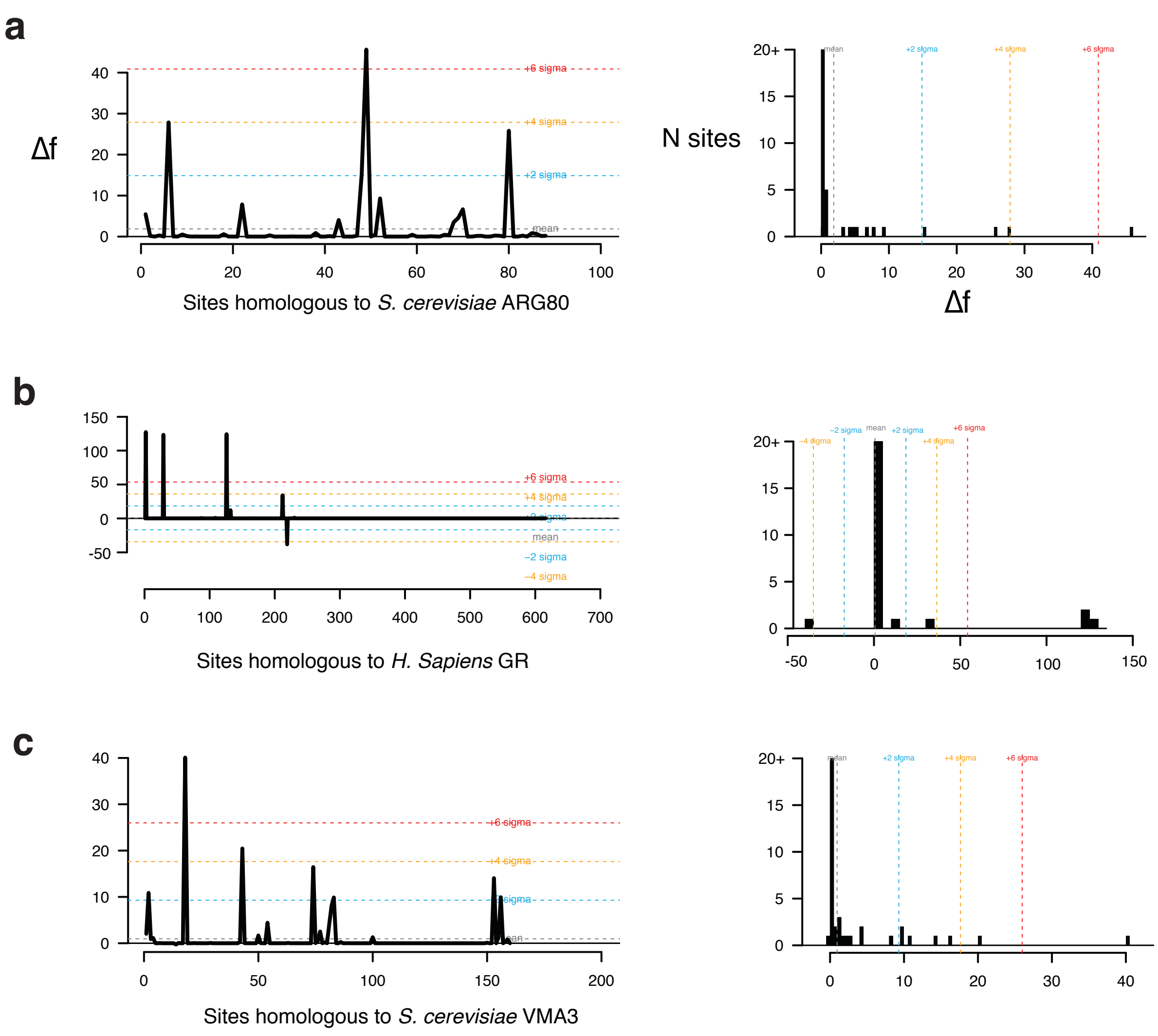

**Supplemental Figure 6. Function-shift scores on three branches with no previously-known functional change.** (a) The Δf scores on the branch leading from Anc.Arg80 in the Mcm1 protein family to the ancestor of *S. cerevisiae* and *Z. rouxii* are plotted for every amino acid site homologous to the *S. cerevisiae* ARG80 sequence. The histogram, on the right side, illustrates the same data binned into groups by Δf score, with bars expressing the number of sites counted in each bin. The Δf scores are similarly shown for mutations in other protein families, on the phylogenetic branches leading to (b) mammalian glucocorticoid receptors in the steroid hormone family, and to (c) Saccharomyces yeast species in the V-ATPase proteolipid subunit family. Bars in the histograms are colored according to their divergence from the mean Δf score. Blue is +/-2 standard deviation, orange is +/-4 s.d., and red is +/- 6 s.d.

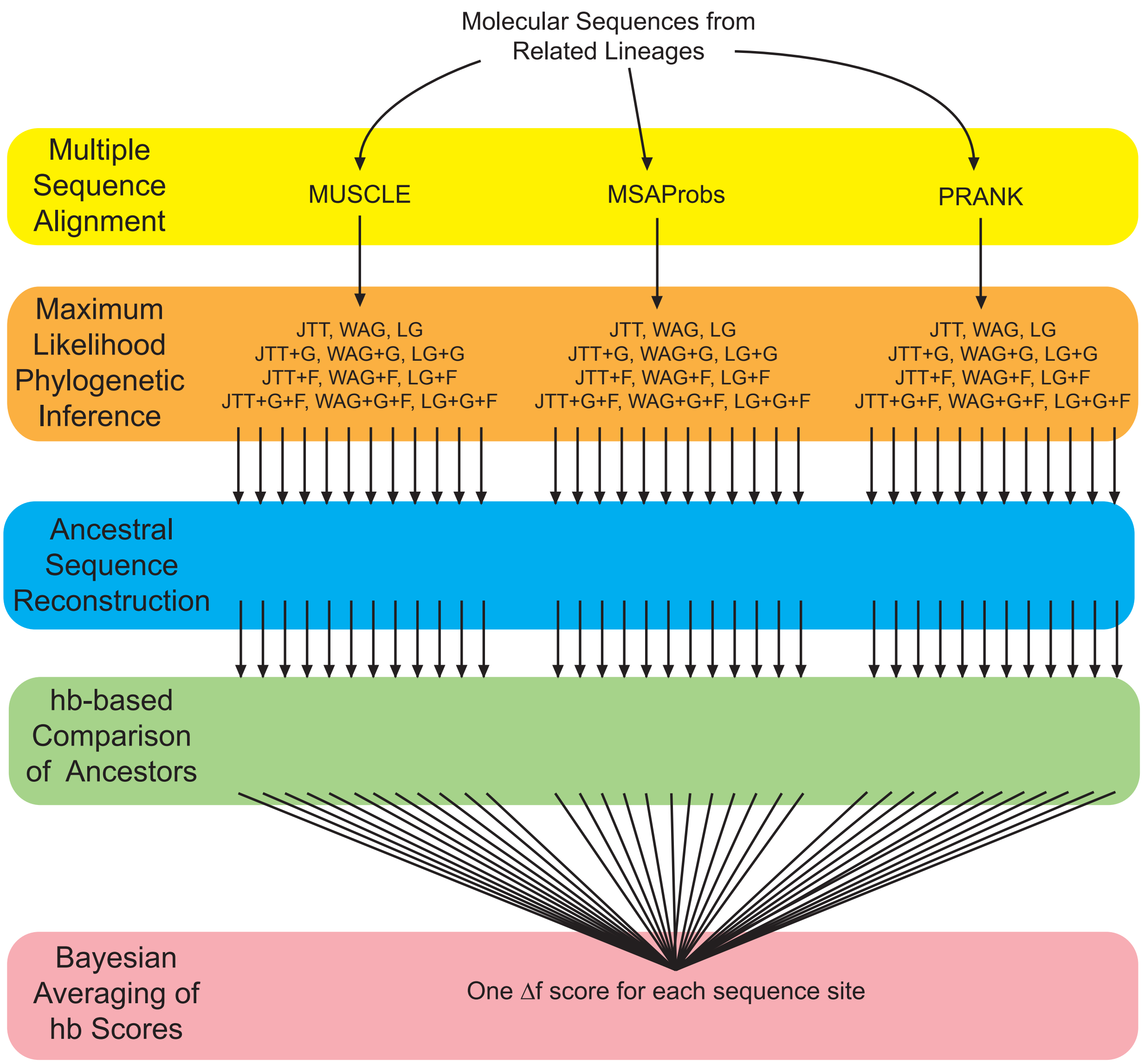

**Supplemental Figure 7. Schematic of software pipeline to integrate alignment and phylogenetic uncertainty.** See Methods and Materials for more description. The pipeline begins with amino acid sequences from a single protein family. The sequences are aligned three different multiple sequence alignment algorithms. A maximum likelihood phylogeny is then inferred for each alignment, using twelve different evolutionary models, to create twelve different phylogenies. On every phylogeny, ancestral sequences are reconstructed and then the $\Delta f$ method is used to compare the ancestors before and after the protein functional shift of interest. Finally, the $\Delta f$ scores from every phylogeny are averaged into a single set of $\Delta f$ scores.